\documentclass[12pt]{article}
\usepackage{amsmath,amssymb}
\usepackage{latexsym,graphicx,epsfig,wrapfig}

\newcommand{\be}{\begin{equation}}
\newcommand{\ee}{\end{equation}}
\newcommand{\bes}{\begin{subequations}}
\newcommand{\ees}{\end{subequations}}
\newcommand{\bea}{\begin{eqnarray}}
\newcommand{\eea}{\end{eqnarray}}
\newcommand{\ba}{$\begin{array}}

\newcommand{\ea}{\end{array}$}

\newcommand{\bear}{\begin{equation}\begin{array}}
\newcommand{\eear}[1]{\end{array}\label{#1}\end{equation}}

\newcommand{\bm}{\boldmath}
\newcommand{\fr}[2]{\dfrac{{ #1}}{{ #2}}}

\newcommand{\fn}[1]{\footnote{{\sf #1}}}

\newcommand{\sqr}{{1\over\sqrt{2}}}

\newsavebox{\fmbox}
\newenvironment{fmpage}[1]
{\begin{lrbox}{\fmbox}\begin{minipage}{#1}}
{\end{minipage}\end{lrbox}\fbox{\usebox{\fmbox}}}

{\end{list}}
\newcounter{enumct}

\allowdisplaybreaks 

\title {\bf \bm Tree-level unitarity constraints
in the most general 2HDM}

\author{\em I.F. Ginzburg$^1$\thanks{email: ginzburg@math.nsc.ru} ,
I.P. Ivanov$^{1,2}$ \thanks{email: igivanov@cs.infn.it}\\
{\small $^1$ Sobolev Institute of Mathematics, Novosibirsk, Russia}\\
{\small $^2$ INFN, Cosenza, Italy}}

\begin{document}

\maketitle

\abstract{ We obtain tree-level unitarity constraints for the most
general Two Higgs Doublet Model (2HDM) with explicit
$CP$-violation.  We briefly discuss correspondence between
possible violation of tree level unitarity limitation and physical
content of the theory.} \vspace{1cm}

The Electroweak Symmetry Breaking in the Standard Model is
described usually with the Higgs mechanism. In its simplest
variant, an initial Higgs field is an isodoublet of scalar fields
with weak isospin $\vec{\sigma}$. The simplest extension of the
Higgs sector known as two-Higgs-doublet model (2HDM) consists in
introducing two Higgs weak isodoublets of scalar fields $\phi_1$
and $\phi_2$ with hypercharge $Y=+1$ (for a review, see
\cite{Hunter} and~\cite{futurereview}).

We consider the Higgs potential of 2HDM in the form
 \bear{c}
V=\dfrac{\lambda_1}{2}(\phi_1^\dagger\phi_1)^2
+\dfrac{\lambda_2}{2}(\phi_2^\dagger\phi_2)^2
+\lambda_3(\phi_1^\dagger\phi_1) (\phi_2^\dagger\phi_2)
+\lambda_4(\phi_1^\dagger\phi_2) (\phi_2^\dagger\phi_1) \\
+\left\{\dfrac{1}{2}\lambda_5(\phi_1^\dagger\phi_2)^2+
\left[\lambda_6(\phi_1^\dagger\phi_1)+\lambda_7
(\phi_2^\dagger\phi_2)\right](\phi_1^\dagger\phi_2) +{\rm
h.c.}\right\}  \\
-\dfrac{1}{2}\left\{m_{11}^2(\phi_1^\dagger\phi_1) +\left[m_{12}^2
(\phi_1^\dagger\phi_2)+ (m_{12}^2)^* (\phi_2^\dagger\phi_1)
\right] +m_{22}^2(\phi_2^\dagger\phi_2)\right\}\,.
 \end{array}\label{higgspotential}
 \end{equation}

Here, $\lambda_{1-4}$, $m_{11}^2$ and $m_{22}^2$ are real (due to
hermiticity of the potential), while $\lambda_{5-7}$ and $m_{12}$
are, in general, complex.

This potential with real coefficients describes the theory
without CP violation in the Higgs sector while complex
values of some coefficients here make CP violation in Higgs
sector possible (a more detailed discussion of many points
here and references see in ref.~\cite{futurereview}).

$\bullet$ The crucial role in the 2HDM is played by the discrete
$Z_2$-symmetry, i.e. symmetry under transformation
\begin{equation}        \label{Eq:Z2-symmetry}
(\phi_1 \leftrightarrow\phi_1,\phi_2 \leftrightarrow-\phi_2) \quad
\text{or} \quad (\phi_1 \leftrightarrow-\phi_1,
\phi_2\leftrightarrow\phi_2).
\end{equation}
This symmetry forbids ($\phi_1,\,\phi_2$) mixing.

With this symmetry, the CP violation in the Higgs sector is
forbidden and Flavor Changing Neutral Currents (FCNC) are
unnatural. In the "realistic"\ theory this $Z_2$ symmetry is
violated.

Potential (\ref{higgspotential}) contains the $m_{12}^2$ term, of
dimension two, which softly violates the $Z_2$ symmetry. Soft
violation implies that $Z_2$ symmetry is broken near the mass
shell, and is restored at small distances $\ll 1/M_i$, where $M_i$
are masses of Higgs particles. The $\lambda_6$ and $\lambda_7$
terms lead to hard violation of the $Z_2$ symmetry.

We classify the Higgs--Higgs states according to the value of
their $Z_2$-parity: $\phi_i\phi_i$, $\phi_i\phi_i^*$, etc. will be
called the $Z_2$-even states, and $\phi_1\phi_2$,
$\phi_1\phi_2^*$, etc. will be called the $Z_2$-odd states.

$\bullet$ The Higgs--Higgs scattering matrix at high enough energy
at the tree level contain only $s$--wave amplitudes; it is described
by the quartic part of the potential only. The {\it tree level
unitarity constraints} require that the eigenvalues of this
scattering matrix be less than the unitarity limit.

Since the coefficients of the scattering matrix at high enough
energy are given only by parameters $\lambda_i$ of the Higgs
potential, the tree-level unitarity constraints are written as
limitations on parameters $\lambda_i$ (for example, in the minimal
SM, with one Higgs doublet and
$V=(\lambda/2)(\phi^\dagger\phi-v^2/2)^2$, such unitarity
constraint looks as $\lambda<8\pi$).

Until now, these constraints were considered only for the case
with softly broken $Z_2$ symmetry and  without CP violation, i.e.
with $\lambda_6=\lambda_7=0$ and real $\lambda_5$, $m_{12}^2$
\cite{Akeroyd:2000wc,kanemura93}. These results were extend for
the case with softly broken $Z_2$ symmetry and with CP violation
in \cite{futurereview}. Here we consider most general case. Our
approach is easily applicable to other extended Higgs sectors
discussed in literature.

$\bullet$ After electroweak symmetry breaking (EWSB) all eight
components of two complex isodoublet fields are transformed into
three Goldstone fields (which are transformed to longitudinal
components of gauge bosons $W_L$, $Z_L$), two charged Higgs bosons
$H^\pm$ and three neutral Higgs bosons $h_1$, $h_2$, $h_3$ (which
might happen to have no definite CP parity).

A natural way for derivation of the tree level unitarity
constraints is to construct the scattering matrix for all the
physical Higgs--Higgs (as well as $Z_L h_i$, $W_L W_L$, etc.)
states in the tree approximation at high enough energy (where
threshold effects are inessential) and diagonalize it. This very
way was realized in the first derivation of such constraint in the
Minimal Standard Model \cite{LQT}.

The tree-level unitarity constraints are written for the
scattering matrix as the limitations for its eigenvalues. They can
be obtained in any basis related to the physical basis by a
unitarity transformation. The derivation simplifies in the basis
of the non--physical Higgs fields $\phi_i^\pm$, $\eta_i$, $\xi_i$,
\cite{Akeroyd:2000wc,kanemura93}.

$\bullet$ In the the high-energy scalar-scalar scattering, the
total weak isospin $\vec\sigma$, the total hypercharge $Y$ are
conserved\fn{This classification  looks more {\em natural} than
both the $O(4)$-classification introduced in \cite{LQT}  (in the
minimal SM) and the scheme based on new quantum numbers $C$, $G$,
and $Y_\pi$ introduced in \cite{kanemura93}. }, with possible
values $Y=0$, 2, $-2$ and $\sigma=0$, 1.

To consider $Y=0$ states, note that the isospinors $\phi_a$
(with $a=1,2$ indicating the doublet) can be represented as
columns, while $\phi_a^\dagger$ are rows. Therefore, the
direct product $\phi_{b\beta}\phi_{a\alpha}^\dagger$
represents a $2\times 2$ matrix for every pair of $a$ and
$b$. Since Pauli matrices plus unit matrix form a basis  in
the space of hermitian $2\times 2$ matrices, one can
write\fn{The first term $A_0$ represents isoscalar, while
the second $\vec{A}$ -- isovector.} $\phi_{b \beta}\phi_{a
\alpha}^\dagger = A_0 \cdot \delta_{\beta\alpha}
+\vec{A}\cdot \vec{\tau}_{\beta\alpha}\; $ with
$A_0=(\phi_{a}^\dagger\phi_{b})/2$,
$A_i=(\phi_{a}^\dagger\tau^i\phi_{b})/2$. Therefore,
 \be
 (\phi_{a \alpha}^*\phi_{b \alpha})(\phi_{c\beta}^*\phi_{d\beta})=
 \fr{1}{2}
 \left[(\phi_{a \alpha}^*\phi_{d \alpha})(\phi_{c \beta}^*\phi_{b \beta})
+  \sum_r (\phi_{a \alpha}^*\tau_{\alpha\beta}^r\phi_{d \beta})
(\phi_{c \gamma}^* \tau_{\gamma\delta}^r\phi_{b\delta})
\right]\,.\label{Firz0}
 \ee

The total set of possible initial states with hypercharge $Y=0$ in
our case can be written as scalar products
 \bear{ccc}
 &\underbrace{Z_2\;\; even}&\underbrace{Z_2\;\; odd}\\[2mm]
Y=0,\;\sigma=0:&\sqr(\phi_1^\dagger\phi_1),\;\;\;\sqr(\phi_2^\dagger\phi_2),&
\sqr(\phi_1^\dagger\phi_2),\;\;\;\sqr(\phi_2^\dagger\phi_1)\,,\\[2mm]
Y=0,\;\sigma=1:&\sqr(\phi_1^\dagger\tau^i\phi_1),\;\sqr(\phi_2^\dagger\tau^i\phi_2),&
\sqr(\phi_1^\dagger\tau^i\phi_2),\;\sqr(\phi_2^\dagger\tau^i\phi_1)\,.
 \eear{Y=0states}
where $i = +,z,-$.

To repeat this trick for the hypercharge $Y=2$ states, we
introduce rows $\tilde{\phi}_a=(i\tau_2\phi_a)^T = (n_a,
-\phi_a^+)$ and corresponding columns
$\tilde{\phi}_a^\dagger = (i\tau_2\phi_a)^*$. Note that the
isoscalar form $\tilde{\phi}_{a\alpha}\phi_{b\alpha}$ is
antisymmetric under permutation $a\leftrightarrow b$,
$\tilde{\phi}_{a\alpha}\phi_{b\alpha}=-\tilde{\phi}_{b\alpha}\phi_{a\alpha}$,
while isovector is symmetric under this permutation.

The total set of possible initial states with hypercharge $Y=2$
can be written similarly to \eqref{Y=0states} as
 \bear{ccc}
 &\underbrace{Z_2\;\; even}&\underbrace{Z_2\;\; odd}\\
Y=2,\;\sigma=0:& absent, &
\sqr(\tilde{\phi}_1\phi_2)=-\sqr(\tilde{\phi}_2\phi_1)\,,\\[2mm]
Y=2,\;\sigma=1:&{1 \over 2}(\tilde{\phi}_1\tau^i\phi_1),\;
{1\over 2}(\tilde{\phi}_2\tau^i\phi_2),&
\sqr(\tilde{\phi}_1\tau^i\phi_2)=\sqr(\tilde{\phi}_2\tau^i\phi_1)\,.
 \eear{Y=2states}
The factor $1/2$ for $Z_2$-even case here is due to the
presence of identical particles in the initial state. The
$Z_2$ even states with $Y=2$, $\sigma=0$ are absent (it
follows directly from the Bose-Einstein symmetry of
identical scalars.). The states with $Y=-2$, $\sigma=1$ are
obtained from those for $Y=2$ by charge conjugation.

Now similarly to \eqref{Firz0} one can write  following relation:
 \be
 (\phi_{a\alpha}^*\phi_{b\alpha})(\phi_{c\beta}^*\phi_{d\beta})=
 \fr{1}{2}
 \left[(\phi_{a \alpha}^*\tilde\phi_{c \alpha}^*)(\tilde \phi_{d \beta}\phi_{b \beta})
+  \sum_r (\phi_{a \alpha}^*\tau_{\alpha\beta}^r\tilde\phi_{c \beta}^*)
(\tilde\phi_{d\gamma} \tau_{\gamma\delta}^r\phi_{b\delta})
\right]\,.\label{Firz2}
 \ee

The scattering matrix  for each set of states with given quantum
numbers $Y$ and $\sigma$ \eqref{Y=0states}, \eqref{Y=2states} in
tree approximation is calculated easily from the potential
(\ref{higgspotential})  with the aid of eq-s \eqref{Firz0},
\eqref{Firz2}. The results are presented in \eqref{scmat}. In each
case left upper corner presents scattering matrix for $Z_2$--even
states and right--down corner -- for $Z_2$--odd states,
coefficients $\lambda_6$, $\lambda_7$ describe mixing among these
states.

 \bes\label{scmat}
 \bea
&
 8\pi S_{Y=2,\sigma=1}&=
 \begin{pmatrix}
\lambda_1 & \lambda_{5}&\sqrt{2}\lambda_6\\
 \lambda_5^*&  \lambda_2&\sqrt{2}\lambda_7^*\\
\sqrt{2}\lambda_6^*&\sqrt{2}\lambda_7& \lambda_3+\lambda_4
\end{pmatrix}\,,\label{YS21}\\
 &
 8\pi S_{Y=2,\sigma=0}&= \;\;\;\lambda_3-\lambda_4\,,\label{YS20}\\
 &
  8\pi S_{Y=0,\sigma=1}&=\begin{pmatrix}
 \lambda_1 & \lambda_4&\lambda_6&\lambda_6^*\\
 \lambda_4&  \lambda_2&\lambda_7&\lambda_7^*\\
\lambda_6^*&\lambda_7^*&\lambda_3&\lambda_5^*\\
\lambda_6&\lambda_7&\lambda_5&\lambda_3
 \end{pmatrix}\,,\label{YS01}\\
 &
8\pi S_{Y=0,\sigma=0}&=\begin{pmatrix}
3\lambda_1&2\lambda_3+\lambda_4&3\lambda_6&3\lambda_6^*\\
  2\lambda_3+\lambda_4& 3\lambda_2&3\lambda_7&3\lambda_7^*\\
 3\lambda_6^*&3\lambda_7^*&\lambda_3+2\lambda_4& 3\lambda_{5}^*\\
 3\lambda_6&3\lambda_7&
 3\lambda_5&\lambda_3+2\lambda_4\end{pmatrix}\,.\label{YS00}
 \eea\ees

The unitarity constraint meant that $S<1$, therefore it limit
eigenvalues of the written matrices $\Lambda$ by inequalities
 \be
 |\Lambda|<\fr{1}{8\pi}\,.\label{unitconstr}
 \ee
Unfortunately, explicit expressions for these eigenvalues are very
complex since they should be obtained from equations of 3-rd or
4-th degree.

However it is useful to present result for the case of softly
broken $Z_2$ symmetry with possible CP violation (obtained in
\cite{GIvunitold} as generalization of \cite{Akeroyd:2000wc}). In
this case $\lambda_6=\lambda_7=0$, our scattering matrices become
block diagonal and their eigenvalues are calculated easily. Now
all eigenvalues of above scattering matrices can be written as
$\Lambda^{Z_2}_{Y\sigma\pm}$
 \bea
 \Lambda^{even}_{21\pm}&=&\fr{1}{2}\left(\lambda_1+\lambda_2\pm
 \sqrt{(\lambda_1-\lambda_2)^2+4|\lambda_5|^2\,}\;\right),
\quad \Lambda^{odd}_{21}=\lambda_3+\lambda_4\,,\quad
\Lambda^{odd}_{20}=\lambda_3-\lambda_4\,,\nonumber\\
\Lambda^{even}_{01\pm}&=&\fr{1}{2}\left(\lambda_1+\lambda_2\pm
 \sqrt{(\lambda_1-\lambda_2)^2+4\lambda_4^2\,}\;\right),
 \quad \Lambda^{odd}_{01\pm}=\lambda_3\pm|\lambda_5|\,,
 \label{eigen}\\
\Lambda^{even}_{00\pm}&=&\fr{1}{2}\left[3(\lambda_1+\lambda_2)\pm
 \sqrt{9(\lambda_1-\lambda_2)^2+4(2\lambda_3+\lambda_4)^2\,}\;\right),
 \ \ \Lambda^{odd}_{00\pm}=\lambda_3+2\lambda_4\pm
 3|\lambda_5|\,.\nonumber
\eea
The obtained constraints \eqref{unitconstr}, \eqref{eigen} differ
from those obtained in ref.~\cite{Akeroyd:2000wc} only by the
change $\lambda_5 \to|\lambda_5|$.

$\bullet$ To describe the general case with hard violation of
$Z_2$ symmetry, we use following fact:
\begin{center}
\begin{fmpage}{0.86\textwidth}
For a Hermitian matrix ${\cal M}=||M_{ij}||$ with maximal and
minimal eigenvalues $\Lambda_+$ and $\Lambda_-$, respectively, all
diagonal matrix elements $M_{ii}$ lie between these eigenvalues,
 \be
\Lambda_+\ge M_{ii}\ge \Lambda_-\,.\label{inequal}\ee\end{fmpage}
\end{center}
This fact follows from extremal properties of the $n$-dimensional
ellipsoid, that is\linebreak[4]  $\Lambda_-\sum x_i^2\le\sum
M_{ij}x_ix_j\le \Lambda_+\sum x_i^2$. (For $2\times 2$ matrix
proof is evident.)

The diagonalization of scattering matrices, yielding their
eigenvalues,  can be realized in two steps. First, we can
diagonalize corners of these matrices corresponding to fixed
values of the $Z_2$ parity. At this step we obtain scattering
matrices with diagonal elements described by
eq-s~\eqref{eigen}. Therefore, by virtue of \eqref{inequal}
the constraints \eqref{unitconstr}, \eqref{eigen} are
\underline{necessary} conditions for unitarity. They are
enhanced due to $\lambda_6$, $\lambda_7$ terms
describing hard violation of the $Z_2$ symmetry. The
shift of eigenvalues \eqref{eigen} caused by these terms
can be easily calculated in the case of a weak hard
violation of $Z_2$ symmetry\fn{In our opinion, precise
equations for eigenvalues with solutions of equations of
3-rd or 4-th degree have no big sense.} $|\lambda_{6,7}|\ll
\Lambda_{Y,\sigma}^{Z_2}$.

$\bullet$ {\bf Some invariants of reparametrization
transformations}. The 2HDM contains two doublet fields,
$\phi_1 $ and $\phi_2 $, with identical quantum numbers.
Therefore, its most general form should allow for global
transformations which mix these fields and change the
relative phase. Each such transformation generates a new
Lagrangian, with parameters given by parameters of the
incident Lagrangian and parameters of the transformation
(the reparametrization invariance) -- see for details
\cite{futurereview}.

The eigenvalues of scattering matrices \eqref{scmat} are --
by construction -- reparametrization invariant quantities.
In the standard equations for eigenvalues of these matrices
their coefficients are also invariants of reparametrization
transformation\fn{ This set of invariants presents
important subset from huge set considered in
ref.~\cite{invHab}. } since they can be constructed from
the eigenvalues. Each $n\times n$ matrix $S_{Y\sigma}$ in
\eqref{scmat} generates $n$ invariant polynomials of
$\lambda_i$: one of the first, one of the second, etc.
order. A convenient choice of these invariants is
$Tr\{S_{Y\sigma}^k\}$ for $k=1,\dots,n$.
($Det\{S_{Y\sigma}\}$ can be also used instead of
$Tr\{S_{Y\sigma}^n\}$.) It gives 12 invariants. Of course,
not all of them are independent. For example, among the
four invariants linear in $\lambda_i$, which we denote
$I_{Y\sigma}\equiv 8\pi Tr\{S_{Y\sigma}\}$ according to
their $Y$ and $\sigma$,
 \bear{l}
I_{21}= \lambda_1+\lambda_2+\lambda_3+\lambda_4\,,\\
I_{20}=\lambda_3-\lambda_4\,,\\
I_{01}=\lambda_1+\lambda_2+2\lambda_3\,,\\
I_{00}=3(\lambda_1+\lambda_2)+2\lambda_3+4\lambda_4\\
 \eear{invariants}
only two are linearly independent. This is related to the
fact that there exist only two independent linear
combination of parameters $\lambda_i$ of the general
quartic 2HDM potential, corresponding to two scalars of the
$SU(2)\times U(1)$ group, describing reparametrization
symmetry \cite{iv2005}.

$\bullet$ {\bf Limitations for masses}. Note that in
the considered tree approximation the masses of Higgs
bosons are composed from quantities $\lambda_i $ and
quantity $\nu \propto Re(\overline{m}_{12}^2)/v_1v_2$ (the
quantity $m_{12}^2$ in a special rephasing gauge different
from that used above -- see \cite{futurereview} and
\cite{masses} for details). Since parameter $m_{12}^2$ does
not enter the quartic interactions, the above unitarity
constraints, generally, do not set any limitation on masses
of observable Higgs bosons, which was explicitly noted in
\cite{kanemura93}. Reasonable limitations on these masses
can be obtained for some specific values of $\nu$. For
example, for reasonably small value of $\nu$ one can have
the lightest Higgs boson mass of about 120 GeV and the
masses of other Higgs bosons can be up to about 600 GeV
without violation of tree-level unitarity
\cite{futurereview}. At large $\nu$, masses of all Higgs
bosons except the lightest one can be very large without
violation of unitarity constraint.

$\bullet$ {\bf Unitarity constraints and strong
interaction in Higgs sector}. Let us discuss briefly some
new features, which are brought up by the situation with
unitarity constraints in 2HDM.

The unitarity constraints were obtained first \cite{LQT} in the
minimal SM. In this model, the Higgs boson mass
$M_H=v\sqrt{\lambda}$, and its width $\Gamma$ (given mainly by
decay to longitudinal components of gauge bosons $W_L$, $Z_L$)
grows as $\Gamma\propto M_H^3$. The unitarity limit corresponds to
the case when $\Gamma_H\approx M_H$, so that the physical Higgs
boson disappears, the strong interaction in the Higgs sector is
realized as strong interaction of longitudinal components of gauge
bosons $W_L$, $Z_L$ at $\sqrt{s}> v\sqrt{\lambda}\gtrsim
v\sqrt{8\pi}\approx 1.2$ TeV. Therefore, if $\lambda$ exceeds the
tree-level unitarity limitation, the discussion in terms of
observable Higgs particle becomes meaningless, and a new physical
picture for the Electroweak Symmetry Breaking in SM arises.

Such type of correspondence among the tree level unitarity limit,
realization of the Higgs field as more or less narrow resonance
and a possible strong $W_LW_L$ and $Z_LZ_L$ interaction, can
generally be violated in the 2HDM if values of $\lambda_i$ differ
from each other essentially. Large number of degrees of freedom of
2HDM generates situations when some of Higgs bosons of this theory
are "normal"\ more or less narrow scalars (whose properties can be
estimated perturbatively), while the other scalars and (or) $W_L$,
$Z_L$ interact strongly at sufficiently high energy. It can happen
that some of the latter can be realized as physical particles,
while the other disappear from particle spectrum like Higgs boson
in SM with large $\lambda$. In such cases the unitarity
constraints work in different way for different {\it physical}
channels. The list of possibilities will be studied elsewhere.

$\bullet$ The scheme proposed above can be readily
exploited in the study of some other multi-Higgs models.
For example, the model with two Higgs doublets plus one
Higgs singlet ($\sigma=0$) with $Y=0$ can be described by
potential \eqref{higgspotential} plus additional terms. In
this case scattering matrices $S_{Y=2,\sigma}$,
$S_{Y=0,\sigma=1}$ have the form \eqref{YS21}-\eqref{YS01}
while scattering matrix $S_{Y=0,\sigma=0}$ is obtained from
that written in \eqref{YS00} by adding of one column and
one row. Besides, only one new scattering matrix
$S_{Y=1,\sigma=1/2}$ appears in this case.\\

We are thankful to M. Krawczyk and V.G.~Serbo for valuable
comments. This work was supported by grants RFBR
05-02-16211 and NSh-2339.2003.2.  The work of IPI was
supported in part by the INFN Fellowship,

\end{document}